\documentclass{article}
\usepackage{spconf}
\usepackage{amsmath}
\usepackage{graphicx}
\usepackage{amssymb}
\usepackage{multirow}
\usepackage{threeparttable}
\usepackage{cite}
\usepackage{booktabs}
\usepackage{makecell}
\usepackage{diagbox}
\usepackage{color}
\usepackage{tcolorbox}
\usepackage[colorlinks=true, linkcolor=blue, urlcolor=blue, citecolor=blue]{hyperref}

\title{SALM: Spatial Audio Language Model with Structured Embeddings for Understanding and Editing}

\name{Jinbo Hu$^{1,2}$, Yin Cao$^{3}$, Ming Wu$^{1}$, Zhenbo Luo$^{2}$, Jun Yang$^{1,4}$}
\address{$^{1}$Institute of Acoustics, Chinese Academy of Sciences, China \quad
$^{2}$MiLM Plus, Xiaomi Inc., China \\
$^{3}$Xi’an Jiaotong Liverpool University, China \quad
$^{4}$University of Chinese Academy of Sciences, China}

\begin{document}
\ninept

\maketitle

\begin{abstract}
Spatial audio understanding is essential for accurately perceiving and interpreting acoustic environments. However, existing audio-language models exhibit limitations in processing spatial audio and perceiving spatial acoustic scenes. To address this gap, we propose the Spatial Audio Language Model (SALM), a novel framework that bridges spatial audio and language through multi-modal contrastive learning. SALM integrates a text encoder with a dual-branch audio encoder that decomposes spatial sound into semantic and spatial components via structured audio embeddings. Key features of SALM include seamless alignment between spatial audio and natural language, both separate and joint extraction of spatial and semantic representations, zero-shot direction classification, and flexible support for spatial audio editing. Experimental results demonstrate that SALM effectively captures and aligns cross-modal representations, yielding well-structured audio embeddings. Furthermore, SALM enables advanced editing capabilities, such as modifying directional audio using text-based embeddings. 
\end{abstract}
\section{Introduction}
Humans naturally interpret spatial acoustic scenes and describe them using language. For instance, the instruction ``When the doorbell in the back rings, open the door" is easily understood. This is because humans can semantically interpret the audio signal ``ringing doorbell", comprehend the spatial concept ``back", and integrate these elements to align semantic meaning with spatial information. Analogously, a machine must also bridge linguistic information with spatial audio to accurately identify, localize, and respond to sounds based on given conditions. 

\begin{figure}[t]
    \centering
    \includegraphics[width=\linewidth]{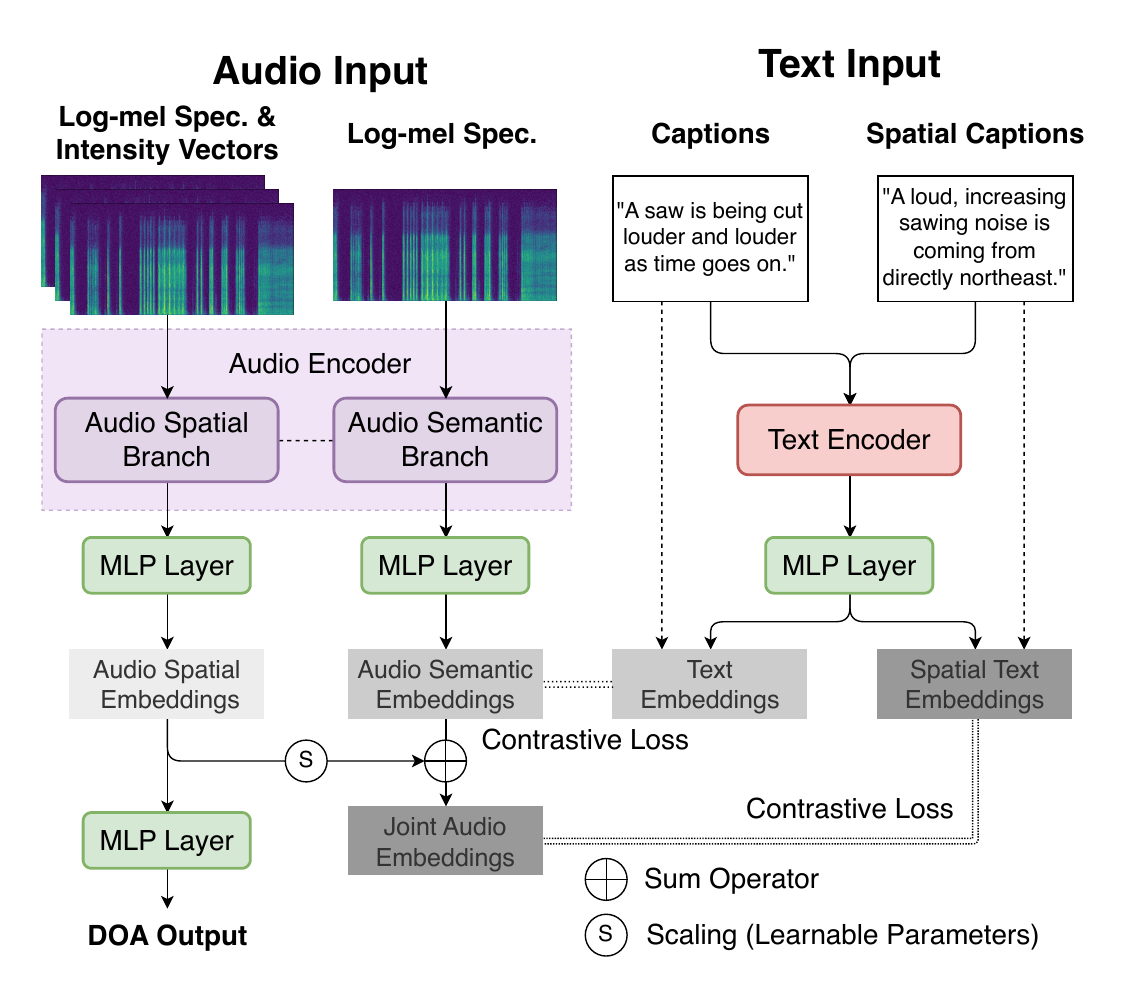}
    \caption{The network architecture of the Spatial Audio Language Model (SALM), comprising a text encoder and a dual-branch audio encoder. The dotted line in the Audio Encoder denotes the learnable parameters connecting two audio branches.}
    \label{fig: sCLAP}
\end{figure}

Audio-language models (ALMs), such as Contrastive Language-Audio Pretraining (CLAP) \cite{ms-clap} and LAION-CLAP \cite{laion-clap}, utilize audio and text encoders along with contrastive learning to establish a correspondence between the two modalities in a shared latent representation space. These models demonstrate exceptional performance in zero-shot audio classification and cross-modal retrieval, exhibiting strong generalization to previously unseen audio recordings and textual descriptions. Building on this paradigm, subsequent studies \cite{pengi, LTU, qwen2_audio, AF3} have incorporated auditory perception into large language models (LLMs), extending their applicability to tasks such as audio question answering and audio captioning. However, most existing models are trained primarily on monophonic audio datasets, which restricts their capacity to represent and leverage the spatial characteristics of sound sources. 

On the other hand, sound event localization and detection (SELD) models \cite{shimada2021accdoa, multiaccdoa, cao2020event, cao2021, hu2022track, wang2023four, nguyen2021salsa, pseldnets, hu2023selective, mff_einv2} focus on jointly detecting sound event categories and estimating the direction of arrival (DOA) of sources in spatial audio. Existing learning-based SELD methods primarily adopt the Activity-coupled Cartesian DOA (ACCDOA) \cite{shimada2021accdoa, multiaccdoa} approach and Event-Independent Network V2 (EINV2) \cite{cao2020event, cao2021, hu2022track} approaches. ACCDOA combines sound event detection (SED) and DOA estimation tasks into a single output by embedding activity information of sound events into Cartesian DOA vectors. In contrast, EINV2 introduces a decoupling mechanism to separate these two subtasks through two relatively independent sub-networks. Building on these methods, our prior work proposed Pre-trained SELD Networks (PSELDNets) \cite{pseldnets}, which serve as foundation models for SELD and demonstrate the benefits of pre-training for adaptation to diverse real-world scenarios. Despite their effectiveness, SELD systems remain constrained to pre-defined sound categories and cannot describe the sound source with natural language. Consequently, neither ALMs nor SELD frameworks can establish an alignment between spatial audio and natural language.

To address these limitations, recent studies \cite{BAT, sudarsanam2025towards, spatial_audio_understanding, text_queried_localization} have begun exploring audio-based LLMs that integrate spatial sound perception into natural language understanding. Most of these approaches are designed for specific tasks-such as spatial sound question answering and reasoning, spatial speech localization and recognition, or text-queried target sound event localization—often framed within a question–answering paradigm. In contrast to such task-specific approaches, Devnani et al. \cite{ELSA} introduced Embeddings for Language and Spatial Audio (ELSA), a CLAP-like model that applies multi-modal contrastive learning on spatially augmented audio–text pairs. By doing so, ELSA learns task-agnostic representations that align spatial audio with corresponding textual descriptions.

In this work, we build upon existing ALMs and SELD models to develop a multi-modal framework. We introduce the Spatial Audio-Language Model (SALM), which aligns spatial audio representations with text representations, as illustrated in Figure \ref{fig: sCLAP}. Inspired by ELSA \cite{ELSA} and our previously proposed PSELDNets \cite{pseldnets}, SALM integrates a text encoder with a dual-branch audio encoder and is trained on synthetic spatial audio-language pairs through contrastive learning. The audio encoder consists of semantic and spatial branches, enabling spatial audio to be decomposed into complementary components that can also be fused into a joint representation through a simple weighted summation. We evaluate SALM through spatial audio-text retrieval and zero-shot direction classification. Experimental results demonstrate that SALM effectively captures and aligns cross-modal embeddings while producing structured audio representations. Notably, SALM enables spatial audio editing by manipulating the structured audio embeddings using text embeddings derived from textural descriptions. For example, arbitrary spatial audio directions can be generated by replacing {Audio Spatial Embeddings} (in Figure \ref{fig: sCLAP}) with text embeddings corresponding to directional descriptions.

\section{Method}
\subsection{Paired Spatial Audio-Language Datasets}

Training a model capable of interpreting spatial audio through natural language requires datasets containing spatial audio annotated with textual spatial descriptions, e.g., ``The bell on the left rings". Popular audio captioning datasets, including AudioCaps \cite{audiocaps} and Clotho \cite{clotho}, provide annotations of textural descriptions for sound event samples from AudioSet \cite{gemmeke2017audio} and Freesound \cite{freesound}. However, these datasets contain only monophonic audio signals and lack explicit spatial information about sound sources. Furthermore, to the best of our knowledge, no publicly available dataset currently provides paired spatial audio–language annotations \cite{ELSA}. To overcome this limitation, we construct a paired dataset for spatial audio-language by using the Clotho and AudioCaps datasets through two pipelines: spatial audio simulation and spatial caption generation.

Following the procedure for generating synthetic spatial sound event samples in PSELDNets \cite{pseldnets}, we synthesize first-order Ambisonics (FOA) audio samples by convolving monophonic audio signals with simulated FOA-format spatial room impulse responses (SRIRs). FOA is an array-agnostic representation that offers flexibility and adaptability across diverse microphone array configurations. The simulated SRIRs are generated by simulating diverse shoe-box rooms, where each room is parameterized by its geometrical dimensions and frequency-dependent absorption coefficients assigned to surfaces \cite{pseldnets}. A virtual microphone is placed at the geometric center of each simulated room to capture SRIRs from various locations within the room. To further enhance realism, we also synthesize spatialized sound events using measured SRIRs from the TAU-SRIR DB \cite{srir-db}, enabling evaluation in real acoustic environments.

The spatial caption generation pipeline enriches textual descriptions by incorporating explicit spatial attributes. In this process, numerical spatial parameters—such as azimuth angles—are first converted into natural language descriptions corresponding to directional classes defined at $45^\circ$ intervals, yielding eight distinct directions (e.g., ``south" or ``northeast"). The original caption, together with its spatial annotation, is then provided to LLaMA 3.2-3B \cite{llama} with an inference temperature of 0.2. This step ensures the generation of spatially enriched captions that precisely describe both the sound events and their spatial positions. An example of the original and generated spatial captions can be found in Fig. \ref{fig: sCLAP}. The prompt used for this process is as follows:

\begin{tcolorbox}[colback=black!5!white, colframe=black!55!white, 
                  fonttitle=\bfseries, fontupper=\footnotesize]
The sound: $<${original caption}$>$ is coming from the $<${direction}$>$. Rephrase the sentence in English to concisely describe the sound detail and the direction of its source. 
\end{tcolorbox}

\begin{table*}[t]
    \centering
    \caption{The performance of spatial retrieval and localization on evaluation sets of sAudioCaps and sClotho. $(\cdot)$ denotes the adopted loss functions. $\mathcal{L}_{\text{CL}}$, $\mathcal{L}_{\text{sCL}}$, $\mathcal{L}_{\text{DOA}}$, referring to retrieval loss, spatial retrieval loss, and DOA loss, can be found in Eq. \ref{eq: loss}.}
    \resizebox{\textwidth}{!}{
        \begin{tabular}{c|ccc|ccc|c|ccc|ccc|c}
        \toprule
        \multicolumn{1}{c|}{\multirow{3}{*}{Model}} & \multicolumn{7}{c|}{sClotho} & \multicolumn{7}{c}{sAudioCaps} \\ \cmidrule(lr){2-8}\cmidrule(lr){9-15}
        & \multicolumn{3}{c|}{Text-to-Audio} & \multicolumn{3}{c|}{Audio-to-Text} & \multirow{2}{*}{\makecell[c]{Local. \\ Error}} & \multicolumn{3}{c|}{Text-to-Audio} & \multicolumn{3}{c|}{Audio-to-Text} & \multirow{2}{*}{\makecell[c]{Local. \\ Error }} \\
        & R@1 & R@5 & R@10 & R@1 & R@5 & R@10 & & R@1 & R@5 & R@10 & R@1 & R@5 & R@10 & \\
        \midrule
        LAION-CLAP ($\mathcal{L}_{\text{sCL}}$) & 2.3\% & 8.8\% & 14.5\% & 2.2\% & 9.0\% & 14.9\% & - & 4.4\% & 18.3\% & 29.0\% & 5.3\% & 20.8\% & 32.4\% & - \\
        SALM-s ($\mathcal{L}_{\text{sCL}}$, $\mathcal{L}_{\text{DOA}}$) & 7.9\% & 24.4\% & 36.1\% & 7.8\% & 22.8\% & 34.7\% & $4.2^\circ$ & 18.5\% & 48.9\% & 63.6\% & 22.6\% & 54.1\% & 67.9\% & $3.1^\circ$ \\
        SALM ($\mathcal{L}_{\text{sCL}}$, $\mathcal{L}_{\text{DOA}}$) & 9.1\% & 28.3\% & 40.5\% & 9.6\% & 28.2\% & 40.6\% & $1.8^\circ$ & 19.6\% & 51.1\% & 65.6\% & 23.7\% & 55.4\% & 68.8\% & $1.3^\circ$ \\
        SALM ($\mathcal{L}_{\text{sCL}}$, $\mathcal{L}_{\text{CL}}$, $\mathcal{L}_{\text{DOA}}$) & \textbf{10.5\%} & \textbf{30.3\%} & \textbf{43.3\%} & \textbf{10.4\%} & \textbf{31.0\%} & \textbf{45.6\%} & $\mathbf{1.6^\circ}$ & \textbf{22.8\%} & \textbf{56.7\%} & \textbf{69.4\%} & \textbf{31.4\%} & \textbf{62.4\%} & \textbf{76.1\%} & $\mathbf{1.2^\circ}$ \\
        \bottomrule
        \end{tabular}
    }
    \label{tab: res}
\end{table*}

\subsection{Model Architecture}
The Spatial Audio-Language Model (SALM) integrates a text encoder and an audio encoder, as illustrated in Fig. \ref{fig: sCLAP}. The text encoder follows the design of LAION-CLAP \cite{laion-clap} and is based on RoBERTa \cite{roberta}, a general-purpose language model built upon the bidirectional transformer architecture \cite{transformer}. The pre-trained checkpoint from the LAION-CLAP text encoder is used to initialize the text encoder. 

The audio encoder adopts the EINV2 variant from PSELDNets \cite{pseldnets} and consists of two complementary branches: an audio semantic branch and an audio spatial branch. The semantic branch processes inputs from the omnidirectional channel of FOA-format signals, while the spatial branch utilizes all FOA channels. This configuration enables two branches to capture the semantic and spatial information of spatial audio independently. To encourage information exchange, the branches are connected through a soft-parameter sharing strategy \cite{cao2021}, implemented via multiple sets of trainable parameters (illustrated by the dotted lines between the two branches in Fig. \ref{fig: sCLAP}). The audio encoder decomposes spatial audio into {Audio Semantic Embeddings} and {Audio Spatial Embeddings}, which can subsequently be merged into {Joint Audio Embeddings} using the weighted sum operation. Both branches adopt the HTSAT \cite{htsat} architecture, a Swin-Transformer-based \cite{swinTransformer} model. For initialization, the Audio Semantic Branch utilizes the pre-trained checkpoint from the LAION-CLAP audio encoder, while the Audio Spatial Branch is initialized with the pre-trained checkpoint of the DOA branch in the EINV2 variant from PSELDNets. 

The Audio Spatial Branch, Audio Semantic Branch, and Text Encoder produce 768-dimensional embeddings. To unify these embeddings, three independent two-layer multi-layer perceptions (MLPs) are applied to each output. These MLPs project 768-dimensional embeddings into 512-dimensional embeddings, ensuring that the final output embeddings from all encoders are dimensionally consistent.

Following LAION-CLAP \cite{laion-clap} and ELSA \cite{ELSA}, our model is trained to learn aligned representations using a batched contrastive loss. This loss encourages the alignment of embeddings from the same sample across different modalities while penalizing the alignment of embeddings from distinct samples. Moreover, given that the simulated rooms used for synthesizing spatial audio are parameterized and the source locations of measured SRIRs in the TAU-SRIR DB are explicitly annotated, we can obtain precise spatial labels for sound sources. To leverage these spatial labels, DOA loss is introduced into the training process. Specifically, the produced 512-dimensional Audio Spatial Embeddings are fed into another two-layer MLP designed to predict the Cartesian DOA of sound sources within the 3D sound scene. The overall loss is defined as:
\begin{equation}
    \mathcal{L}_{\mathrm{SALM}}=\frac{1}{2}(\mathcal{L}_{\text{CL}}+\mathcal{L}_{\text{sCL }})+\mathcal{L}_{\text{DOA}},
    \label{eq: loss}
\end{equation}
where $\mathcal{L}_{\text{CL}}$ is the contrastive loss between {Audio Semantic Embeddings} and {Text Embeddings}, $\mathcal{L}_{\text{sCL}}$ is the contrastive loss between {Joint Audio Embeddings} and {Spatial Text Embeddings}, and $\mathcal{L}_{\text{DOA}}$ is the cosine distance between the predicted and target DOAs for sound sources. The loss functions contain an aggregated item ($\mathcal{L}_{\text{sCL}}$) and two decoupled items ($\mathcal{L}_{\text{CL}}$ and $\mathcal{L}_{\text{DOA}}$), explicitly designed to enable the model to enhance semantic understanding and spatial perception jointly and independently.

\subsection{Embedding Structure}
\label{sec: embedding_structure}
The SALM text encoder generates embeddings from input textual descriptions. Specifically, {Text Embeddings} are derived from the original captions, while {Spatial Text Embeddings} are produced from the spatially enriched captions, as shown in the Text Input component of Fig. \ref{fig: sCLAP}. 

The SALM audio encoder utilizes two decoupled branches to generate output audio embeddings, i.e., {Audio Spatial Embeddings} ($E_\mathtt{ASp}$) and {Audio Semantic Embeddings} ($E_\mathtt{ASe}$). These two embeddings are subsequently merged into {Joint Audio Embeddings} ($E_\mathtt{JA}$) using the following operation:
\begin{equation}
    E_\mathtt{JA} = E_\mathtt{ASe} + \mathbf{s} \odot E_\mathtt{ASp},
\end{equation}
where $\mathbf{s}$ is a set of learnable weights with 512 dimensions and $\odot$ denotes the Hadmard product. The formulation allows the model to adaptively integrate semantic and spatial information, thereby yielding a comprehensive spatial audio representation. 

Furthermore, SALM enables spatial audio editing by manipulating structured audio embeddings with text embeddings derived from target descriptions. For example,  arbitrary spatial directions can be imposed by replacing the Audio Spatial Embeddings $E_\mathtt{ASp}$ with text embeddings corresponding to directional descriptions, such as ``The sound is coming from the southwest." We denote these embeddings as $E_\mathtt{TDi}$. The process is formalized as follows:
\begin{equation}
    \tilde{E_{\mathtt{JA}}} = E_\mathtt{ASe} + \mathbf{s} \odot \frac{\lVert E_\mathtt{ASp}\rVert \cdot E_\mathtt{TDi}}{\lVert E_\mathtt{TDi}\rVert},
\end{equation}
where $\lVert\cdot\rVert$ denotes the L2 norm, which is applied to normalize $E_\mathtt{TDi}$, ensuring that its scale is consistent with that of $E_\mathtt{ASp}$. The approach enables flexible control of spatial audio directionality directly from textual descriptions.

\section{Experiments}
\subsection{Experimental Setups}
\label{sec: exp_setup}
The synthetic spatial datasets are built upon Clotho \cite{clotho} and AudioCaps \cite{audiocaps}, containing 5,929 and 50,956 audio clips, respectively. Each audio clip is spatially augmented by simulating three distinct source locations within the same room, enabling the model to observe identical audio from multiple spatial perspectives in each epoch. This design integrates both semantic and spatial cues, providing a more comprehensive basis for evaluation. In total, we synthesize 17,787 spatial audio clips from Clotho (approximately 111 hours), referred to as {sClotho}, and 152,868 audio clips using the AudioCaps dataset (approximately 419 hours), denoted as {sAudioCaps}. To facilitate evaluation under real acoustic scenes, we further construct equivalent-scale datasets, sAudioCaps-R and sClotho-R, by incorporating measured SRIRs from TAU-SRIR DB \cite{srir-db}.

Following LAION-CLAP \cite{laion-clap} and ELSA \cite{ELSA}, we evaluate spatial retrieval performance by computing cross-modal audio-to-text and text-to-audio matching across all samples. Retrieval performance is measured using the cosine similarity between embeddings from the two modalities. Specifically, we report recall at rank 1 (R@1), rank 5 (R@5), and rank 10 (R@10), which indicate whether the correct match is retrieved within the top 1, 5, or 10 candidates, respectively. In addition, to assess spatial accuracy, we calculate the angular distance between the predicted and ground-truth DOAs, thereby quantifying localization error.

The sampling rate of synthetic spatial audio samples is 24 kHz. We extract the 64-dimensional log mel spectrograms and intensity vectors \cite{pseldnets} from four-channel FOA signals with a Hanning window of 1,024 points and a hop size of 240 as input audio features. Each input segment has a fixed duration of 10 seconds during both training and inference. For clips shorter than 10 seconds, the audio is repeated and then zero-padded to meet the required length, while longer clips are randomly cropped to 10-second segments. Training is conducted with a batch size of 64 using the Adam optimizer. The learning rate schedule includes a linear warm-up phase over the first three epochs, followed by cosine decay over the subsequent seven epochs, with a basic learning rate of $10^{-4}$.

\subsection{Spatial Semantic Retrieval}

\begin{table*}[t]
    \centering
    \caption{The performance of spatial retrieval and localization on evaluation sets of sAudioCaps-R and sClotho-R.}
    \resizebox{\textwidth}{!}{
        \begin{tabular}{c|ccc|ccc|c|ccc|ccc|c}
        \toprule
        \multicolumn{1}{c|}{\multirow{3}{*}{Training Datasets}} & \multicolumn{7}{c|}{sClotho-R} & \multicolumn{7}{c}{sAudioCaps-R} \\ \cmidrule(lr){2-8}\cmidrule(lr){9-15}
        & \multicolumn{3}{c|}{Text-to-Audio} & \multicolumn{3}{c|}{Audio-to-Text} & \multirow{2}{*}{\makecell[c]{Local. \\ Error}} & \multicolumn{3}{c|}{Text-to-Audio} & \multicolumn{3}{c|}{Audio-to-Text} & \multirow{2}{*}{\makecell[c]{Local. \\ Error }} \\
        & R@1 & R@5 & R@10 & R@1 & R@5 & R@10 & & R@1 & R@5 & R@10 & R@1 & R@5 & R@10 & \\
        \midrule
        sAudioCaps, sClotho & 6.3\% & 18.2\% & 26.3\% & 6.1\% & 17.7\% & 25.1\% & $12.7^\circ$ & 13.5\% & 36.3\% & 48.7\% & 13.7\% & 33.5\% & 44.4\% & $12.6^\circ$ \\
        + {sAudioCaps-R, sClotho-R} & 7.2\% & 22.2\% & 32.2\% & 7.8\% & 21.9\% & 30.7\% & $7.6^\circ$ & 16.4\% & 42.6\% & 55.9\% & 21.4\% & 46.4\% & 56.7\% & $7.5^\circ$ \\
        \bottomrule
        \end{tabular}
        }
    \label{tab: res_colrir}
\end{table*}

We conduct experiments to evaluate the spatial audio understanding capabilities of various models and loss functions, using spatial retrieval metrics and localization errors, as shown in Table \ref{tab: res}. 

SALM-s is a variant of the SALM and only utilizes the Audio Spatial Branch, in contrast to the dual-branch architecture of the full model. A key distinction between LAION-CLAP \cite{laion-clap} and SALM-s lies in their handling of input audio channels. While LAION-CLAP utilizes only one omni-direction channel of FOA audio signals, SALM-s incorporates all channels. Experimental results show that SALM-s consistently outperforms LAION-CLAP across all metrics, primarily because LAION-CLAP lacks spatial information in its input and thus relies solely on semantic content for audio–text matching.

Extending SALM-s with an additional Audio Semantic Branch further improves performance, particularly in localization accuracy, where localization error is reduced by approximately 60\%, as illustrated in the middle two rows of Table \ref{tab: res}. This demonstrates SALM's ability to effectively align and integrate spatial and semantic information. Additionally, incorporating $\mathcal{L}_{\text{CL}}$ into optimization yields further performance improvement across all metrics. While $\mathcal{L}_{\text{sCL}}$ and $\mathcal{L}_{\text{DOA}}$ guide cross-modal alignment and DOA estimation, respectively, $\mathcal{L}_{\text{CL}}$ functions as a regularizer, ensuring consistency between the Audio Semantic Embeddings and the Text Embeddings.

Table \ref{tab: res_colrir} presents the performance on sAudioCaps-R and sClotho-R, which are synthesized using measured SRIRs in real acoustic environments. The results demonstrate that SALM, when trained on sAudioCaps and sClotho with simulated SRIRs, generalizes effectively to these real-scene datasets. Moreover, augmenting training with additional synthetic data generated from measured SRIRs further improves the performance across all evaluation metrics. Importantly, the measured SRIRs used for the training and evaluation originate from distinct recording environments, underscoring the robustness of SALM’s generalization. Consistent with PSELDNets \cite{pseldnets}, SALM exhibits strong transferability to real-world scenarios.

\subsection{Zero-shot Direction Classification}

\label{sec: zsl_doa}
\begin{table}[t]
    \centering
    \caption{Accuracy of zero-shot direction classification (8-class).}
    \resizebox{0.6\linewidth}{!}{
        \begin{tabular}{c|cc}
            \toprule
             Feature & sClotho & sAudioCaps \\
             \midrule
             Audio Sem. Embed. &8.2\% & 12.2\% \\
             Audio Spat. Embed. & 99.9\% & 100.0\% \\
             Joint Audio Embed. & 99.9\% & 100.0\% \\
             \bottomrule
        \end{tabular}
    }
    \label{tab: res_zsl_doa}
\end{table}

We generate captions through the template ``\texttt{The sound is coming from the <direction>}", where \texttt{direction} represents one of the eight-direction descriptions, such as ``south" or ``northeast". Text embeddings for these captions are extracted using the pre-trained SALM Text Encoder and compared with audio embeddings from spatial audio samples using cosine similarity. A match is classified as correct if the audio embedding is the closest to the text embeddings derived from the corresponding direction descriptions. 

Zero-shot direction classification accuracy is shown in Table \ref{tab: res_zsl_doa}. The results indicate that Audio Semantic Embeddings fail to align effectively with text embeddings associated with directional descriptions, likely due to the absence of spatial information. In contrast, both {Audio Spatial Embeddings} and {Joint Audio Embeddings} demonstrate strong alignment with the corresponding text embeddings. These observations further suggest that the spatial and semantic information in spatial audio can be effectively decoupled and then reintegrated, yielding comprehensive and robust audio representations.

\subsection{Spatial Audio Editing}

\begin{figure}
    \centering
    \includegraphics[width=\linewidth]{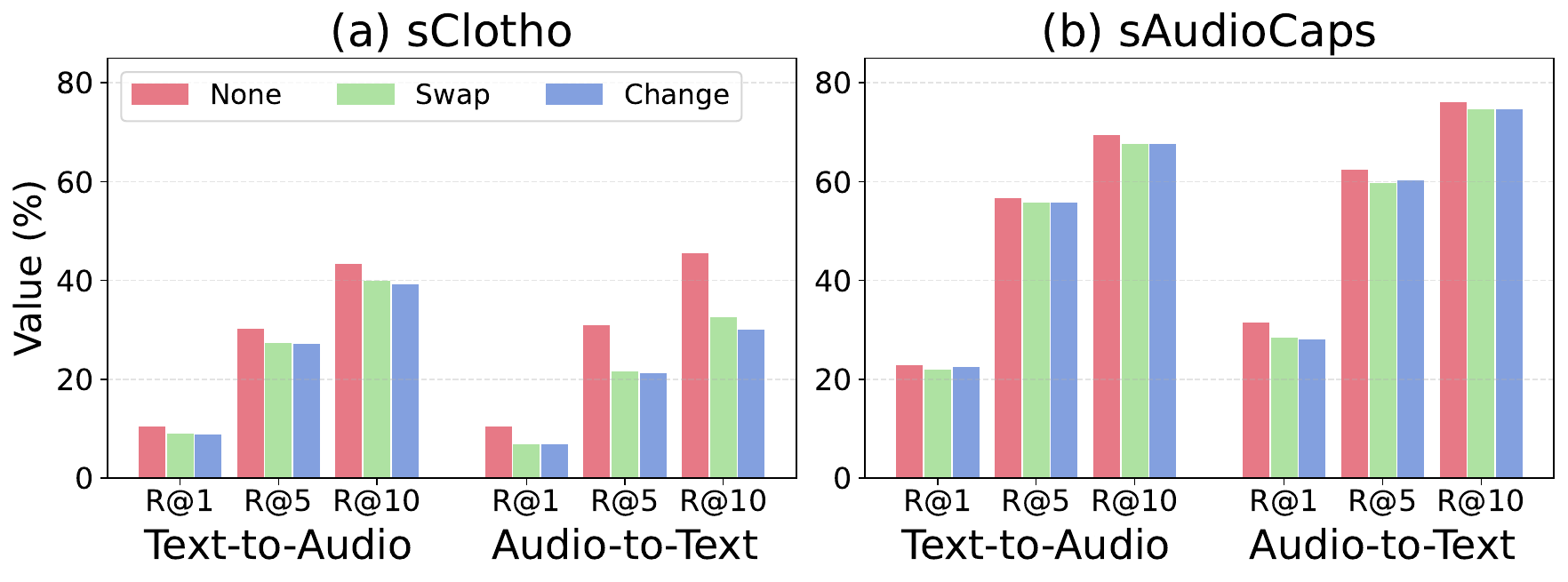}
    \caption{The spatial retrieval performance while editing spatial audio. ``None", ``Swap", and ``Change" refer to no operation, direction-invariant editing, and direction modification, respectively.}
    \label{fig: res_editing}
\end{figure}

We manipulate the direction of spatial audio by modifying the {Audio Spatial Embeddings} item (details in Sec. \ref{sec: embedding_structure}) and present the results in Fig. \ref{fig: res_editing}. The term ``None" indicates that no operation in audio embeddings, ``Swap" refers to replacing {Audio Spatial Embeddings} with text embeddings generated from the templates described in Sec. \ref{sec: embedding_structure} and \ref{sec: zsl_doa}, while preserving the original directions of {Joint Audio Embeddings}. ``Change" denotes altering the sound direction by substituting the Audio Spatial Embeddings with text embeddings corresponding to different directional descriptions. Overall, we find that ``Swap" and ``Change" operations minimally impact the spatial-semantic information of sound. Experimental results show that {Joint Audio Embeddings} are well-structured and can be edited without substantially compromising the underlying semantics of the sound.

\section{Conclusion}
We presented the Spatial Audio Language Model (SALM), a framework designed to align spatial audio with natural language through multi-modal contrastive learning on paired spatial audio-text datasets. SALM integrates a text encoder and a dual-branch audio encoder that decomposes spatial audio into semantic and spatial representations via structured audio embeddings. Experimental results show that SALM effectively captures and aligns cross-modal features, as validated through spatial audio–text retrieval. Furthermore, SALM generates well-structured and flexible audio embeddings, enabling tasks such as zero-shot direction classification and spatial audio editing. Specifically, spatial audio editing is achieved by manipulating structured audio embeddings, allowing directional alterations guided by text-derived embeddings. These findings highlight SALM’s potential as a foundation for future applications in spatial audio understanding, reasoning, and generation.

\bibliographystyle{IEEEbib}
\bibliography{refs}

\end{document}